# Phishing Website Detection Using a Combined Model of ANN and LSTM

Muhammad Shoaib Farooq, Hina jabbar

School of System and Technology, University of Management and Technology, Lahore, 54000, Pakistan

Corresponding author: Muhammad Shoaib Farooq (e-mail: shoaib.farooq@umt.edu.pk).

**ABSTRACT** In this digital era, our lives highly depend on the internet and worldwide technology. Wide usage of technology and platforms of communication makes our lives better and easier, but on the other side it carries out some security issues and cruel activities, phishing is one activity of these cruel activities. It is a type of cybercrime, which has the purpose of stealing the personal information of the computer user, and enterprises, which carry out fake websites that are the copy of the original websites. The attackers used personal information like account IDs, passwords, and usernames for the purpose of some fraudulent activities against the user of the computer. To overcome this problem researchers focused on the machine learning and deep learning approaches. In our study, we are going to use machine learning and deep learning models to identify the fake web pages on the secondary dataset. The data set is comprised of 48 attributes and 10,000 instances, of which 5,000 instances consist of the data set of phishing websites, and 5,000 instances dataset consist of legitimate websites. To investigate the fake websites five models used machine learning which are decision tree, k-nearest neighbor, naive byes, logistic regression, SVM, and four models of deep learning Artificial Neural Network, LSTM, and a purposed hybrid model ANN-LSTM for comparing and evaluating the performances of the model's different matrices of measurement are used including accuracy, f1 score, recall, and precision Purposed hybrid model ANN-LSTM performs better than other models with 98 percent accuracy and k-nearest Neighbor performs poorly with an accuracy of 74 percent because the lowest number of k=100 using the large numbers of k is computationally expensive to get the result.

**INDEX TERMS** Cybercrime, Webpages Legitimate, Illegitimate, Machine Learning, Deep Learning, LSTM, ANN,

## I. INTRODUCTION

Instant growth in the technologies, source of communications, a platform of e-commerce, the vast management information system, every computer has the ability to store the information and share this information with others are the major reasons for the cyber-crime [1][2]. Cybercrime is a hot issue in the perspective of security problems in this era for all individual users and organizations they are informed users and informed organizations, even though alertness and informed and trained users are necessary bus impossible to be a victim of these types of attacks and scams [3][4]. Phishing is a type of cybercrime [5] [6]. The attackers of these types of attacks create fake web pages similar to the original website [7] [8]. They use many ways for attacks like URL, E-mail, internet protocol, messaging, and other Tele communication ways [9] [10]. The approach that is used commonly for a phishing scam is a spoofing web that copies the original website [11]. It is critical for the individual users and the organizations to deduct the fake websites due to the strong structure of the attackers, and classify whether the web page is fake or not fake [8]. Personal information of the computer user and use this information for the purpose of fraud against the use [12]. It was noted by a report that Phishing attacks have grown day by day in recent years. Anti-Phishing Working Group (APWG) notes that phishing attacks have increased in recent years, fig1 shows the number of phishing attacks increased by 10 percent quarterly and 40 percent yearly. According to this report phishing attacks occur 10 percent in the first quarter of 2017 which have grown increasingly 10 percent to 100 percent at the end of 2020. Phishing attacks have caused damage to many organizations and the economy, in the last quarter of 2020 [11].

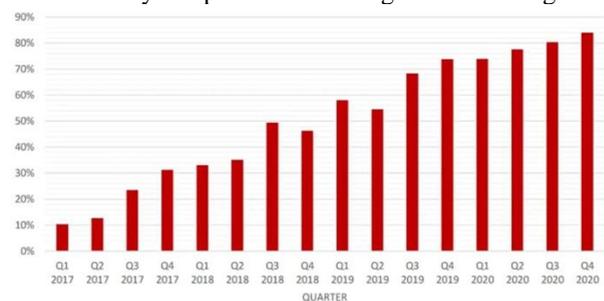

FIGURE 1. **PHISHING ATTACKS YEARLY REPORT**

Many ways have been used to solve the security issues and classify the phishing web pages to the original web pages [13]. Phishing crime increases day by day [14]. The automated signature-based technique signature based used to detect fake websites which is not able to identify some fake websites [15]]. For example, a method that is defined as black listed method detection approach, technique lists the website using URLs that are classified as fake sites and legitimate sites. The main drawback of the blacklist method is that, not able to identify zero hour phishing attack [16] [17][18]. To solve these problems, now concentrate on the techniques namely ML which is able to classify the fake websites as legitimate websites and provide high accuracy [19].

We focused on deducting the fake websites in real-time by examining the uniform resource locators of the web pages using different machines of ML and DL approaches [10] [20]. We are going to apply five models of ML and four DL algorithms and one is a proposed hybrid model of deep learning models which are LSTM-ANN.48 features are used to detect fake web pages like IP addresses and URLs using abnormal characters.

The major contributions of this paper are providing the best architecture that provides high accuracy for the prediction of phishing websites to extract useful categorized accuracy for the prediction of phishing attacks and evaluation of different machine learning and deep learning algorithms on phishing dataset which is comprised of the phishing websites and legitimate website.

## II. RELATED WORK

The machine learning approach is also used for anti- phishing techniques established on different features from different sources [21]. On the basis of features the model identifies websites that are phishing and legitimate [3]. The dataset consists of 72000 uniform resource allocator compresses of some significant and worthless expressions and words like dot, and hyphen space. Using these characters also detaches the levels of the domains. Identifying the characters, expressions, and special words. They are formed with others with others characters [15]. This paper deployed six machine learning classifiers which are NB, RF, and KNN. Where the value of the N is 3 and DT. They are used to detect fake web pages and compare the performances of all classifiers [22]. According to the results, RF performs well and provides 97 percent accuracy this result is satisfactory and good for deducting the fake web pages identifying the words, which are formed with a random character. For the NB, feature engineering improves the performance of the classifiers. The researchers worked on 6000 URL-based datasets in which 10 more features were extracted to identify between phishing and legitimate websites. They use 8 machine learning classifiers. it is seen that the multi-layer perceptron (MLP) classifier achieves the highest level of accuracy of 85.41 percent as compared to the other classifiers. The researchers proposed a comparative analysis over different datasets and estimated the performance of machine learning classifiers [5]. The ANN achieves 97 percent accuracy. The objective of this study is to propose a hybrid rule-based solution by combining 6 classifiers to detect fake web pages. Black list, host, identity, lexical, content, visual similarity, identity similarity, and behavioral methods 37 features are extracted. Classifiers include CART, SVM, KNN, MLP, and CNN. The results show that CNN achieves 97 percent accuracy[23]. Using third-party services, the available website content data are major cause of phishing attacks [1]. Deep learning classifiers, e.g., DNN, CNN, and RNN are used for identifying fake webpages. Researchers did 72 percent identification by using the FS method. They use the phish tank dataset. Researchers focus on detecting fake websites using the DL framework [16]. The main purpose of this paper is to create 2 different types of fake web features: original and interaction features. The model DBN presented is called Deep Belief Networks. In the testing phase from internet service provider IP flows to detect the fake webpages. The DBN achieves 90 percent accuracy [24]. Some researchers create a framework called phish safe established on the FS method [25]. There are a total of fourteen features based on fake and original web pages. SVM performs best. Phishing attacks are rapidly increasing nowadays in the world. The attacker's aim is to get the personal information of the user and ethically wrong use on the legitimate website [8]. Many researchers claim that many people lose their expensive things and money.

Through machine learning and deep learning techniques, we can detect and prevent unwanted attacks. The researchers review different approaches like content, heuristic, and fuzzy-based approaches [26]. For IOT service provider fake website is a big problem nowadays. Several techniques are used for the detection of fake web pages [27]. Machine learning and deep learning classifiers NB, DT, Neural Networks, and RF are used for detection. The researchers focus on detection using IOT devices. SVM performs best [28]. Using zero-day website attacks, researchers propose a new solution. They incorporate white lists and visual similarity methods. Researchers use a computer vision technique which is called a SURF detector. SURF works for feature extraction from legitimate and fake websites [29]. In this paper, to deduct the phishing websites use the DL algorithms and to classify the current previous work into numerous types. ML and DL algorithms includes, CNN, NLP, DNN, RNN.

## III. DESCRIPTION OF DATASET

A diverse dataset containing samples of both legitimate and phishing websites to train and evaluate the model. Using publicly available datasets like the Phishing Websites Data Set from Kaggle. The dataset comprises 48 features as shown in table I and consists on 10000 instances 5000 phishing websites and 5000 legitimate websites which are not fake. Basically to fit the model training dataset used which 70% of our dataset, and the remaining 30% dataset is used to calculate the model effectiveness and prediction of the binary classification[30]. In x train represent the features for the training data, in x test signifies features for testing data, in y train represents predicted variables for training data, and in y test symbolizes non predicted variable for testing data an enhanced feature extraction method is engaged by leveraging the browser automation framework, which is more detailed and strongly associated with the analyzing method based on regular languages [31]. Data preprocessing involves:

- Attainment the dataset
- Introducing data sets
- Discovery missing data
- Encoding categorical data
- Splitting dataset into training and test set
- Feature scaling

**TABLE I. FEATURES**

| ID | Random String | Sub Domain Level |
|---|---|---|
| Path Level | URL Length | Number Percent |
| No Https | IP Address Query | Domain in Path |
| Https in Hostname | Components | Brand Name |
| Double Slash in Path | Number Sensitive | URL Length RT |
| Sub Domain Level | Words | Click Disabled |
| RT Images Only in | Fake Link in Status Bar | Path Length |
| Form Query Length | Host Name Length | Missing Title |
| Domain Mismatch | Submit Info to Email | Pop Up Window |
| Domain In | Insecure Forms | CLASS LABEL |
| Subdomains | Abnormal Action R | |

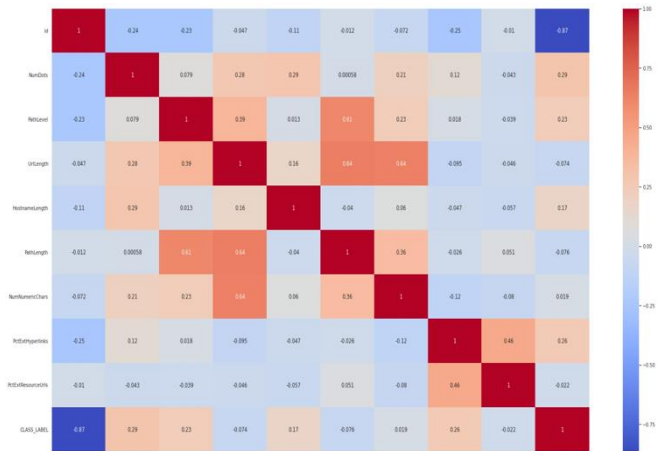

FIGURE 2. **Correlation of Features**

By using correlation of features extract all feature which have high similarities with the class labels.

## IV. METHODOLOGY

Machine learning approaches and deep learning promise the best result for classification problems at this time [30]. using machine learning approach and deep learning for phishing webpage detection is that it can be efficiently assimilated into existing systems of security and workflows. For example, machine learning and deep learning algorithms can be used to automatically scan incoming emails and messages and flag any messages and emails that carry links to phishing web pages. This paper also used an empirical method to investigate the performance of the hybrid model. Now phishing is a classification problem, so machine learning and deep learning classifiers are updated techniques to detect phishing websites [32]. Based on learning models identifying the fraudulent transaction classify quickly, so in our paper methodology uses some supervised classification models where the models train using the train dataset with labels, these models created some output on the basis of input.

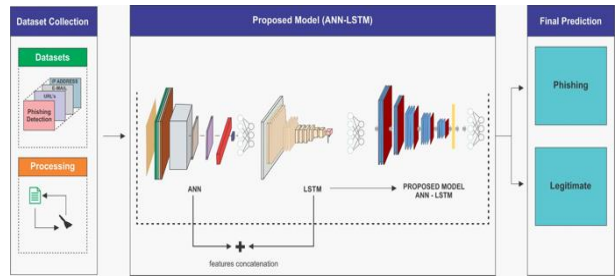

FIGURE 3. **Architecture Diagram**

The major purpose is to integrate two algorithms to raise the performance of the models in that way they perform well in comparison to the algorithms that perform independently in characterization. Recurrent ANNs are able to learn the dependence of order in the issues of forecast and then used in the problem of natural language process applications. LSTM is another deep neural network that we are going to integrate with the ANN model to make a better model for achieving better characterization results over the single LSTM and ANN provide themselves.

*ANN-LSTM*

ANN-LSTM is our proposed empirical model which con- sists of the combination of ANN and LSTM models as shown in Figure. 3 in order to achieve the best performance of both models and the benefits of both models. ANN and LSTM are shown to be excellent performances in overcoming the classification problem, task recognition and the detection problem [33]. Using this empirical method for detecting phishing websites is promising. As a result, by using deep learning models effectively, we were encouraged to find the solution to detect fake websites [34]. In this paper, we used an empirical model which is the combination of ANN, and LSTM to generate the best results for the detection of phishing websites. Therefore, the objective of this paper is to achieve the best accuracy to classify the URLs whether they are fake or legitimate by using the combined model. The suggested model is very helpful and useful for deep learning-based approaches and classification approaches in the field of security and IS (information security) and cybersecurity [35]. It is helpful for stopping financial losses and cybercrimes by classifying phishing websites as legitimate websites. Our work contributes to the effectiveness of using ANN, LSTM, and ANN-LSTM. These are the following points that explain the contribution of this proposed work: A well-planned examination of all methods that are used currently to identify the URLs to identify the fake and legitimate websites. Analysis of the three models LSTM, ANN and combined ANN-LSTM of deep learning approach to predict the fake and legitimate websites. Compare and evaluate the approach LSTM, ANN, and ANN- LSTM architecture.

*B. LSTM*

LSTM is another type of recurrent neural network which is an adaptive model [17]. It is that type of recurrent neural network that has a memory cell [36]. In addition, in the internal state, every neuron switches to the conservative neuron. Each layer of the LSTM architecture consists of memory blocks, which are respectively connected to the other memory blocks, connections of the memory blocks found in each block of each layer of LSTM. As a result, in an LSTM every cell has its own input gate which controls all inputs across the outside of the cell and it also determines the data is overlooked which is inside as well in the output gate that describes the ability of the inner state to view outside from the cells [37]. LSTM provides an efficient strategy for detecting phishing websites[38]. The workflow of the LSTM model for the problem of classifying URLs starts from loading, and then preprocessing and splitting the data. The model starts with the input layer that uses a vector which has a length of 79 and then the layer which is hidden from the LSTM model which consists of the 128 neurons and an output layer using the sigmoid function.

*C. ARTIFICIAL NEURAL NETWORK*

An Artificial Neural Network is used for classification problems, this model structure consists of artificial neurons, which are consist on the three basic of the layer's sequence Artificial Neurons consist on the three basic layers, Input layer, Hidden layer, and Output layer [39]. In an Artificial Neuron Network receives the input from the input layer adds the bias value on the hidden layer and generates the output on the output layer, The shortest neural network is that perceptron which carries n parameters and has only one neuron, and only one output [40]. The n represents the total number of the attributes in the dataset. Forward proliferation is a method that is used to transmit all data across the neural network, it is all done by using a perceptron. By multiplying the all- input values with weights and by adding the results. Using the activation function controls all weight that is on the input layer, then using the activation function select the attribute whether it would be fire or not when a node fired creates output layer [34]. Then in the non-linear function, we have passed the value of many activations functions that exist for the different problems but we have used to sigmoid function in this paper because we have predicted the probability of the binary classification problem.

$$\sum = X.W = X_1 \cdot W_1 + X_2 \cdot W_2 \ldots + X_n w_n \qquad (1)$$

Where the sigma denotes the sigmoid activation function and whether results, we get is called predictable result which is denoted by the Y.

$$Z = X \cdot W + B \qquad (2)$$

The same calculations are done in each layer which is hidden and, in the output, layers are used to pass the values

$$\overline{Y} = \sigma(Z) = \frac{1}{1+e^{-z}} \qquad (3)$$

## V. EXPERIMENTS AND RESULTS

*A. Environment Setup*

The experiments setup is designed by using different deep learning algorithms, python language and data analytics libraries including Scikit Learn that were used for importing the models, classifiers, and parameters for measuring the performance, KERAS was used for importing deep learning models, NumPy is used for some mathematical operations and pandas were used to read the dataset, and Matplotlib was used for plotting the results. Using these libraries we employed deep learning algorithms namely Artificial Neural Network, LSTM, ANN-LSTM. Colab is an excellent and effective tool for data scientists to execute machine learning models and deep learning projects with cloud storage abilities Colab is basically a cloud-based tool which provide environment that requires no setup

*B. Evaluation Matrices*

For evaluate the performance and quality of the model, different measure and metrics are used, and these metrics are known as performance metrics or measure and evaluation metrics or measures. One of them is Confusion Matrix, the confusion matrix is a measure which is used to describe the process of the classification classifiers for a certain and specific set of unseen data (test data). It assesses the performance of the classification models when they create predictions on test data. Accuracy. It is one of the efficient and significant parameters to describe the accuracy of the binary classification problems. It also describes how efficiently and perfectly the model predicts the correct output Precision: It may be described the number of accurate outputs delivered by the model or out of all positive classes that are predicted properly by the model. Recall: It is well-defined as out of all positive classes, how our model predicted perfectly. The recall measure should be as great as can possible. F1-Score measure If two models have small precision and high recall, it is very problematic which is associate these models. So, for this problem, we use for the solution of this problem the F1-score.

*C. Evaluation of Results*

For evaluation of phishing websites classification performance, we use accuracy recall, precision, and F1 score confusion matrix measures of all classifiers. Recall is the measures the percentage of phishing and legitimate pages that the model manages to detect (models' perfections) which is shown in figure 5. Precision measures the standard to which the phishing detected websites are also phishing. The F1 score is the weighted mean of both precision and recall measures. We use 1 as a fake website and 0 as not fake website on our binary classification problem. Use various Machine Learning classifiers to calculate and evaluate the accuracy, precision, recall, and f1- score of various algorithms which is shown in Table II.

TABLE II. **EXPERIMENTAL RESULT**

| Classifier's | Accuracy | Precision | Recall | F1-Score |
|---|---|---|---|---|
| ANN | 94% | 96% | 96% | 96% |
| Logistic Regression | 97% | 98% | 98% | 98% |
| Decision Tree | 93% | 92% | 92% | 92% |
| Naïve Bayes | 92% | 90% | 90% | 93% |
| KNN | 74% | 82% | 82% | 71% |
| SVM | 95% | 92% | 92% | 92% |
| LSTM | 85% | 86% | 86% | 85% |
| BI-LSTM | 93% | 93% | 93% | 91% |
| ANN-LSTM | 98% | 97% | 97% | 97% |

Table II presents the classifiers outcomes for dataset. The highest accuracy is achieved by ANN-LSTM indicating 98%. While KNN indicated the lowest accuracy (74%). It can be observed that the KNN classifier has very low F1 score (71%). But we observed Logistic Regression also perform well but not better than the combined model.

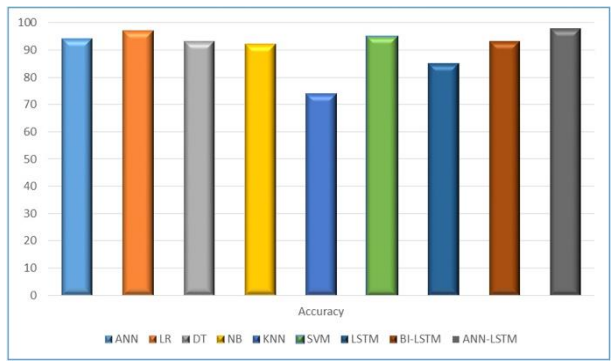

FIGURE 4. **Comparison of accuracy**

FIGURE 4 shows the results of model's performance graphically.

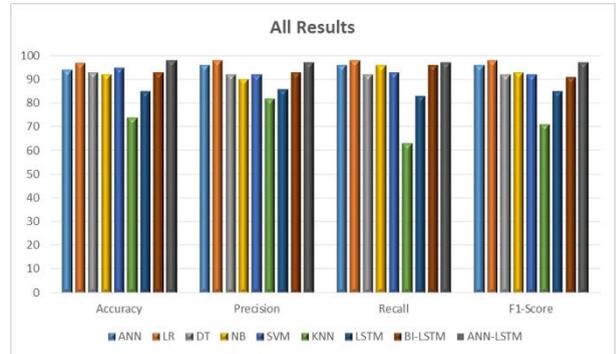

FIGURE 6. **ALL RESULTS.**

We observed in Figure 6 Logistic Regression and ANN also provide high precision, recall and F1 score but we are going to compare overall measures to evaluate the results. After evaluation of all measures of results our proposed model perform better other than all others existing models.

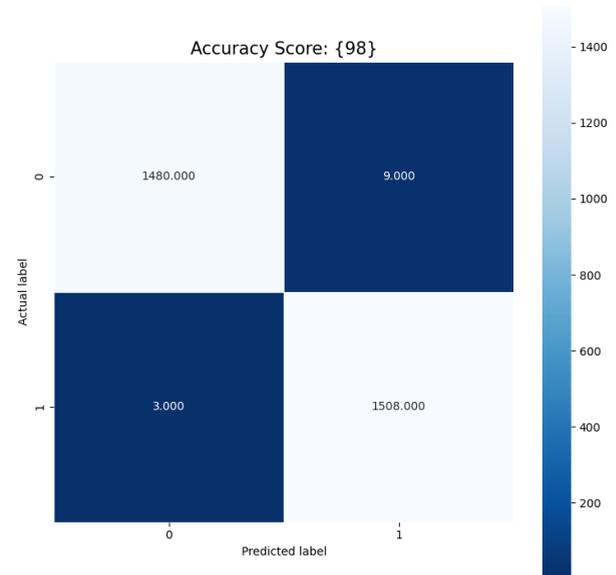

FIGURE 5. **confusion matrix**

The model has predicted as shown in fig 5, 1480 true positive values, phishing websites, which are the actual phishing websites, and 1508 true negative values which are not phishing websites, which are actually not phishing websites. The model has predicted 9 false negative values which are phishing websites, which are actually not phishing website and 3 false negative values which are not phishing, but actually phishing websites. It shows proposed model perform better by evaluating using evaluation matric (confusion matrix)

TABLE IV. **COMPARISION WITH EXISTING MODELS**

| Model's | Accuracy | Approach |
|---|---|---|
| KMR | 89% | Classifier's (SVM-NLP |
| GAN | 90% | Framework |
| SHR | 97% | Classifier's(LSTM-GRU) |
| AMD | 97% | Classifier's(RF-ANN) |
| ATH | 90% | Classifier's(KNN-SVM) |
| ABW | 93% | Classifier's(CNN-LSTM) |
| MKV | 90% | Framework(Protection for ICS) |
| BBR | 97% | Classifier's(RF-NLP) |
| AWM | 97% | Classifier's(SVM-FFA) |
| Proposed Model | 98% | Classifier's(LSTM-ANN) |

TABLE IV shows the comparison of the accuracy combined model of ANN-LSTM with the existing models which proposed by other authors in their researchers. The statics shows our model performed better than the all-other existing approaches by achieving high accuracy (98%).

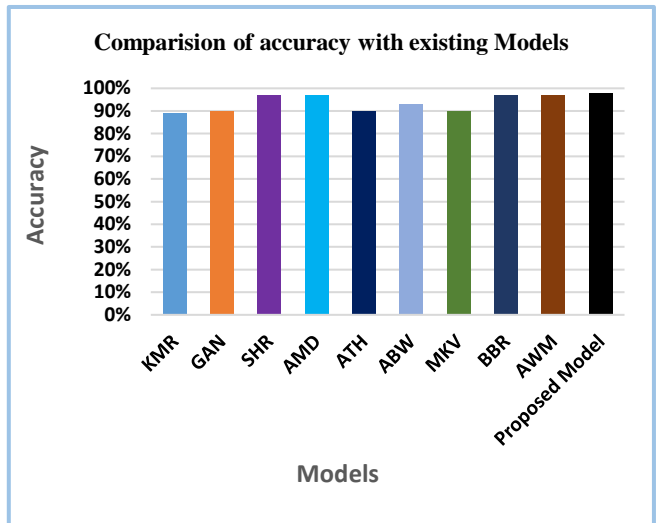

FIGURE 5. **Comparison of Accuracy**

In FIGURE 5 shows graphically the comparison of the existing model's accuracy with our proposed model (LSTM-ANN). Our proposed model provides highest accuracy for detection of phishing websites.

## VI. CONCLUSION

By using machine learning approaches and deep learning approaches can detect fake websites because these websites have some same characteristics. By using automated tools and updated tech- niques easy to detect these phishing websites. We have applied in our research estimated six models on the secondary data set which consists of 48 attributes and 10000 instances, which comprises on 5000 phishing website dataset and 5000 legitimate websites dataset. The inspected five models used machine learning which are decision tree, k-nearest neighbor, naive byes, logistic regression, SVM, and four models of deep learning Artificial Neural Network, LSTM, and a proposed hybrid model LSTM–ANN. Our proposed hybrid empirical method performs better than other models with 98 percent accuracy and k-Nearest Neighbor performs poorly with an accuracy of 74 percent because the lowest number of k=100 using the large numbers of k is computationally expensive to get the result. but when we are considering the accuracy issue, our proposed method is the best performance compared to the other methods. But when considering precision, recall, and F1-Score issues, we see that logistic regression performs best as compared to the other methods. Combining the multiple models will perform better than the individual model it is worthy to declare. Notify that the results of the models encourage future work by adding more features in the dataset, it will help to improve the performance of the classifiers.

The code of experiment and dataset available on GitHub.

https://github.com/hinanaz231/abc